\documentclass[a4paper,fleqn,usenatbib]{mnras}

\usepackage{newtxtext,newtxmath}

\usepackage[T1]{fontenc}
\usepackage{ae,aecompl}

\usepackage{graphicx}	
\usepackage{amsmath}	
\usepackage{amssymb}	

\defcitealias{McQuillan_et_al_2014}{McQ14}

\title[A Kepler Study of Starspot Lifetimes]{A \textit{Kepler} Study of Starspot Lifetimes with Respect to Light Curve Amplitude and Spectral Type}

\author[H. A. C. Giles, A. Collier Cameron \& R. D. Haywood]{
Helen A. C. Giles,$^{1,2}$\thanks{E-mail: Helen.Giles@unige.ch}
Andrew Collier Cameron$^{2}$
and Rapha\"{e}lle D. Haywood$^{3}$
\\
$^{1}$Observatoire de Gen\`{e}ve, Universit\'{e} de Gen\`{e}ve, Chemin des Maillettes 51, Versoix, 1290, Switzerland\\
$^{2}$Centre for Exoplanet Science, SUPA, School of Physics and Astronomy, University of St Andrews, North Haugh, St Andrews, KY16 9SS, UK\\
$^{3}$Harvard-Smithsonian Center for Astrophysics, 60 Garden Street, Cambridge, MA 02138, USA\\
}

\date{Accepted XXX. Received YYY; in original form ZZZ}

\pubyear{2016}

\begin{document}
\label{firstpage}
\pagerange{\pageref{firstpage}--\pageref{lastpage}}
\maketitle

\begin{abstract}
Wide-field high precision photometric surveys such as \textit{Kepler} have produced reams of data suitable for investigating stellar magnetic activity of cooler stars. Starspot activity produces quasi-sinusoidal light curves whose phase and amplitude vary as active regions grow and decay over time.
Here we investigate, firstly, whether there is a correlation between the size of starspots - assumed to be related to the amplitude of the sinusoid - and their decay timescale and, secondly, whether any such correlation depends on the stellar effective temperature.
To determine this, we computed the autocorrelation functions of the light curves of samples of stars from \textit{Kepler} and fitted them with apodised periodic functions. The light curve amplitudes, representing spot size were measured from the root-mean-squared scatter of the normalised light curves. We used a Monte Carlo Markov Chain to measure the periods and decay timescales of the light curves.
The results show a correlation between the decay time of starspots and their inferred size. The decay time also depends strongly on the temperature of the star. Cooler stars have spots that last much longer, in particular for stars with longer rotational periods. This is consistent with current theories of diffusive mechanisms causing starspot decay. We also find that the Sun is not unusually quiet for its spectral type - stars with solar-type rotation periods and temperatures tend to have (comparatively) smaller starspots than stars with mid-G or later spectral types.

\end{abstract}




\section{Introduction}

The \textit{Kepler} mission was designed to search for extrasolar planet transits in stars (within a single field of view) in particular small, Earth-like planets around Sun-like stars \citep{Borucki_et_al_2010,Koch_et_al_2010,Jenkins_et_al_2010}. It has provided insight into planet formation as well as new exoplanet discovery, which also allowed to determine occurrence rates \citep{Howard_et_al_2012,Petigura_Howard_Marcy_2013,Kane_Kopparapu_DomagalGoldman_2014,Burke_et_al_2015,Dressing_Charbonneau_2015,Santerne_et_al_2016a} and further probe the statistics of exoplanet population and system architectures.

\textit{Kepler} has also revolutionised stellar physics. Tens of thousands of stars have 4 years worth of almost continuous, high precision photometry, allowing for a thorough study of stellar brightness modulations across different stellar ages and types. From \textit{Kepler}, fields such as asteroseismology \citep{Bastien_et_al_2013} and differential rotation studies \citep{Reinhold_Reiners_Basri_2013, Aigrain_et_al_2015,Balona_Abedigamba_2016} of main sequence stars have evolved through the study of such a large sample of stars. \citet{McQuillan_et_al_2013,McQuillan_et_al_2014} (hereafter known as \citetalias{McQuillan_et_al_2014}) made the first large-scale surveys of stellar rotation by analysing the autocorrelation functions of stellar light curves.

This unprecedented wealth of high-precision, continuous photometric data for thousands of main-sequence stars has enabled us to take a new look at our own Sun, resulting in comparisons between it and stars which are Sun-like. \citet{Gilliland_et_al_2011} \citep[and pre-\textit{Kepler},][]{Radick_et_al_1998} found that the Sun appears to be unusually inactive when compared to other solar-type stars, but it has since been suggested that this may in fact not be the case \citep{Basri_Walkowicz_Reiners_2013}. This is discussed in \S~\ref{sec:solar}.
In this paper our goal is to discover how Kepler observations can be used to infer the lifetimes of active regions on other stars, and to determine how the lifetime of an active region depends on its size and on the stellar photospheric temperature. 

We define stellar activity, and active regions, in this context as meaning phenomena that introduce surface brightness inhomogeneities, giving rise to apparent flux modulation as the star rotates. Measurements of solar irradiance as a function of wavelength show that bright faculae and dark starspots are the main contributors to solar flux modulation on timescales of order days to weeks \citep{Foukal_Lean_1986}. These modulations have a greater amplitude when the Sun is near the maximum of its 11-year activity cycle. The solar irradiance variations are complex; solar active regions often comprise a bipolar spot group surrounded by an extended facular region of enhanced surface brightness. As an active region crosses the solar disc, the limb brightening of the faculae and foreshortening of the dark spots tends to cause a net initial flux increase. This is followed by a decrease as the spot visibility increases and the facular limb brightening declines \citep{Fligge_Solanki_Unruh_2000}. A similar pattern is seen in Kepler light curves. At times of high activity, the amplitude of variability is often seen to increase with no obvious change in the mean flux level in the Kepler bandpass. Solar irradiance measurements, however, show clearly that the facular flux increase outweighs the dark spot deficit at times of high activity \citep{Lockwood_et_al_2007}.

For the Sun, a range of activity levels have been observed since telescopic records began (from the Maunder Minimum to large-amplitude cycles in the mid-20th century) and there are many differing opinions on what constitutes `typical' solar activity levels \citep{Hanslmeier_et_al_2013,Usoskin_et_al_2016,Inceoglu_et_al_2015,Krivova_Balmaceda_Solanki_2007,Wehrli_Schmutz_Shapiro_2013,Livingston_et_al_2007}. The consensus appears to be that the average level of solar activity lies in between the extremes observed in the past 400 years. For our purpose, we will use the activity levels seen in the last 3 to 4 sunspot cycles as typical levels.

Furthering our understanding of stellar activity is not only important to the stellar community; it is crucial to many other areas of investigations, particularly in the exoplanet society. The presence of starspots and other magnetic active regions can induce quasi-periodic variations over timescales of weeks to months. These activity signatures are seen as major sources of noise in the search for small exoplanets (Earths and super-Earths); spots can lead to wrong planet radius measurements \citep{Barros_et_al_2014}.
The presence of starspots and other magnetically active regions are a real nuisance in RV exoplanet observations. As well as starspots, faculae and granulation produce signals modulated by the star's rotation. They evolve over time, giving rise to quasi-periodic signals with varying amplitudes and phases. This induces RV variations of 1-2 ms$^{-1}$ even in the quietest stars \citep{Isaacson_Fischer_2010}. Stellar noise can conceal and even mimic planetary orbits in RV surveys, and has resulted in many false detections (eg. CoRoT-7d, \citealt{Haywood_et_al_2014}; Alpha Centauri Bb, \citealt{Rajpaul_et_al_2015}; HD166435, \citealt{Queloz_et_al_2001}; HD99492, \citealt{Kane_et_al_2016}; HD200466, \citealt{Carolo_et_al_2014}; TW Hydra, \citealt{Huelamo_et_al_2008}; HD70573, \citealt{Soto_Jenkins_Jones_2015}; HIP13044, \citealt{Jones_Jenkins_2014}; Kapteyn's Star, \citealt{Robertson_Roy_Mahadevan_2015}; Gliese 667d, \citealt{Robertson_Mahadevan_2014}; and GJ 581d \citealt{Robertson_et_al_2014}). It also significantly affects our mass estimates, which are routinely determined from RVs.
A number of methods have been developed to account for activity-induced RV signals and have been quantitatively tested to review their perfomance \citep{Dumusque_2016,Dumusque_et_al_2017,Haywood_et_al_2016,Rajpaul_et_al_2015}.
Therefore, knowing the active region lifetimes can provide significant constraints for models used to determine exoplanet properties, such as mass \citep[see][]{Lopez_et_al_2016}.
Additionally, planet radii and masses are central to theoretical models of planet composition and structure \citep[e.g.][]{Zeng_Sasselov_2013} and are essential to interpreting observations of atmospheres \citep[see][]{Winn_2010}.
When it comes to studying atmospheric transmission spectroscopy of planet atmospheres, un-occulted spots serve to increase the ratio of the area of the planet's silhouette to that of the bright photosphere, making the transit look deeper than it really is. On the other hand, un-occulted faculae have the opposite effect. Since the contrast of both faculae and spots against the quiet photosphere depends on wavelength, particular care has to be taken in the interpretation of atmospheric transmission spectroscopy \citep{Pont_et_al_2007,Oshagh_et_al_2016,Chen_et_al_2017}. As the effects of starspots and suppression of the granular blueshift in faculae are expected to diminish towards longer wavelengths \citep{Marchwinski_et_al_2015}, forthcoming infrared RV spectrometers such as CARMENES \citep{CARMENES} and SPIRou \citep{SPIRou} may help to separate planetary reflex motions from stellar activity signals.
However, until recently only optical spectrometers were reaching the precision needed to determine the masses of super-Earth planets but CARMENES has been achieving $\rm 2ms^{-1}$ which is sufficient for measuring super-Earths \citep{CARMENES_2016} which would therefore suggest that others will be able to perform similarly, according to their specifications. 

Sunspot (and by association, starspot) decay lifetimes have been a point of interest for decades, with many theories for the cause of their decay and what function it follows. Numerical investigations such as those by \citet{Petrovay_MorenoInsertis_1997, Petrovay_DrielGesztelyi_1997, Litvinenko_Wheatland_2015, Litvinenko_Wheatland_2016} indicate that sunspot decay is consistent with a parabolic decay law, where the area of the spots decreases as a quadratic function of time. Observations of the Sun \citep{MorenoInsertis_Vazquez_1988, MartinezPillet_et_al_1993, Petrovay_DrielGesztelyi_1997, Petrovay_et_al_1999, Hathaway_Choudhary_2008} have similarly reflected the same behaviour. This relationship would imply that the main factor in spot decay is granulation, which was first hypothesised by \citet{Simon_Leighton_1964}.
Extrapolating the physics observed to occur on the Sun, only a few attempts have been made to measure starspot decay lifetimes. These studies would allow us to test our theories for sunspot decay on other Sun-like stars. As we cannot resolve the surfaces of others stars directly and at high-resolution like we can for the Sun, their sizes over time have to be inferred from indirect indicators. \citet{Bradshaw_Hartigan_2014, Davenport_Hebb_Hawley_2015, Aigrain_et_al_2015} have recovered the decay lifetime of starspots from both real and simulated \textit{Kepler} data. However, there has not been a large-scale survey of starspot decay lifetimes until now.

In this paper, we determine the starspot lifetimes in a large sample of stars selected to have rotation periods close to 10 days and 20 days. Our technique, based on MCMC parameter estimation, allows us to determine estimates and uncertainties for the stellar rotation period and starspot lifetime of each star. We then investigate how the decay lifetimes relate to extrapolated spot sizes and whether the stellar spectral type has a role in this relationship. In \S~\ref{sec:samples}, we justify the choice of stellar targets. In \S~\ref{sec:methods} we describe our improvements to the method used in \citetalias{McQuillan_et_al_2014} and how the representative measurements for spot sizes are determined. In \S~\ref{sec:results} and \ref{sec:conclusions} we outline and discuss our results and the implications they have for stellar physics and exoplanetary discovery and characterisation. 

\section{Sample Selection}
\label{sec:samples}

Our samples are drawn from the sample of stars analysed by \citetalias{McQuillan_et_al_2014}. They analysed over 34,000 main sequence stars taken from the \textit{Kepler} mission stellar archive at the NASA Exoplanet Archive \citep{Akeson_et_al_2013}. All of the stars in \citetalias{McQuillan_et_al_2014} were less than 6500K in temperature and excluded known eclipsing binaries (EBs) and \textit{Kepler} Objects of Interest (KOIs). \citetalias{McQuillan_et_al_2014} utilised T$_{\rm eff}$ - $\log\,g$ and colour-colour cuts used by \citet{Ciardi_et_al_2011} to select only main sequence stars. The boundary of 6500K was selected by \citetalias{McQuillan_et_al_2014} to ensure that only stars with convective envelopes, that spin down during their lifetime, were included.

To keep computational time to manageable levels, two samples were drawn from the +34,000 \citetalias{McQuillan_et_al_2014} stars based on the measured rotation periods. Sample 1 has a range of periods between 9.5 and 10.5 days, and sample 2 with a range of 19.5 to 20.5 days. This resulted in 1089 and 1155 stars in each respectively. Unlike in \citetalias{McQuillan_et_al_2014} where they used quarters 3-14 from the Exoplanet Archive, quarters 1 to 17 were used here. This was done to extend the temporal span of the light curves as much as possible.

\section{Methods}
\label{sec:methods}

\begin{figure}
	\includegraphics[width=\columnwidth]{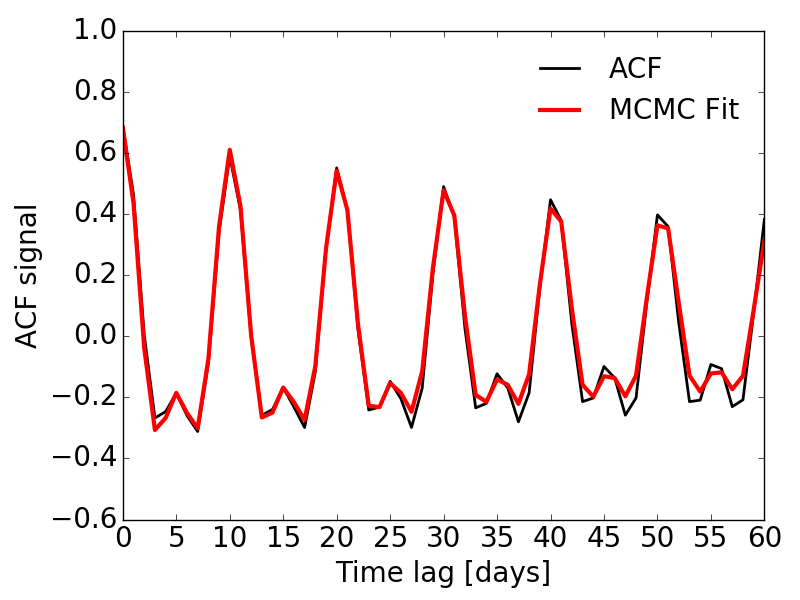}
    \caption{Example of a fitted auto-correlation function for \textit{KIC 8869186} using eq.~\ref{eq:uSHO2}. Selecting the positive time lag half of an ACF, it follows a similar pattern as an underdamped simple harmonic oscillator (uSHO) which has a functional form that can be fitted using Monte Carlo Markov Chains (MCMC).}.
    \label{fig:fitted_acf}
\end{figure}

\subsection{Autocorrelation Function (ACF)}
We created ACFs in the same fashion as \citet{McQuillan_et_al_2013,McQuillan_et_al_2014} who cross-correlated each \textit{Kepler} light curve with itself at a series of discrete timeshifts (time lags). The correlation increases and decreases dependening on the presence of a large dominant starspot. As a light curve can be approximated as sinusoidal in shape \citep{Jeffers_Keller_2009}, a time lag at an integer multiple of the stellar rotation period correlates strongly meaning the first side lobe of an ACF corresponds to the stellar rotation period with further side lobes as harmonics of the period.
The decrease in side lobe amplitude at higher time lags occurs as the light curve gradually varies in amplitude and phase due to starspot formation and decay. Therefore the decay rate of the side lobes describes the decay rate of the starspots. By visual inspection, this appears to be comparable to an exponential decay. With this knowledge, ACFs were fitted with a simple analytical function. This is an improvement on what was reported by \citet{McQuillan_et_al_2013,McQuillan_et_al_2014} as it establishes further parameters of the stellar activity but also determines errors for them.

Many autocorrelation algorithms require the data to be uniformly sampled in time -- \textit{Kepler} data is close to uniformity but has variation in exact observation times and has significant data gaps. Therefore to generate ACFs, the light curves were binned and weighted as described by \citet{Edelson_Krolik_1988}, which has the added advantage of providing error estimates.
Once the ACFs were generated, they were orthogonalised by subtracting the inverse variance-weighted mean, to ensure there were no unwanted correlations between the ACF power and the time lag.

The behaviour of an ACF at zero time lag $\geq0$ days resembles the displacement of an underdamped simple harmonic oscillator (\textit{uSHO}), described by
\begin{equation}
\label{eq:uSHO1}
y(t) = e^{-t/\tau_{AR}} \left( A \cos \left(\frac{2\pi t}{P}\right) \right) + y_0.
\end{equation}

Many ACFs have an additional `interpulse' close to half of the stellar rotation period (Fig.~\ref{fig:fitted_acf}). This corresponds to there being another large but less dominant starspot on the opposite side of the star. Therefore the \textit{uSHO} equation was adapted to include an inter-pulse term,
\begin{equation}
\label{eq:uSHO2}
y(t) = e^{-t/\tau_{AR}} \left( A \cos \left(\frac{2\pi t}{P}\right) + B \cos\left(\frac{4\pi t}{P}\right) + y_0 \right).
\end{equation}
\textit{$\tau_{AR}$} is the decay timescale [days] of the ACF which represents the decay timescale of the dominant starspot. \textit{P} is the stellar rotation period [days$^{-1}$]. (Parameters \textit{A}, \textit{B} and $y_0$ do not represent physical properties of the star, but are needed to fit the \textit{uSHO} equation.) \textit{A} and \textit{B} are the amplitudes of the cosine terms and $y_0$ is the offset of the \textit{uSHO} from $y=0$. The stellar rotation period is taken to be the time lag at which the largest side lobe occurs at and is found by searching for all peaks in the ACF and establishing which is the highest (besides the peak at time lag = 0 days).

\citet{Brewer_Stello_2009} used a damped, stochastically-driven harmonic  oscillator model to emulate the quasi-periodic behaviour of solar p-modes. They also computed the autocorrelation function of the resulting time series, obtaining an expression equivalent to eq.~\ref{eq:uSHO1} above. They used this as the kernel for a gaussian-process regression analysis of the waveform. Because of the $N^3$ computational overhead involved in Gaussian-process regression, the large number of data points in each light curve and the large number of light curves analysed here, we elected instead to perform the parametric fit to the autocorrelation functions, as described by eq.~\ref{eq:uSHO2}.

\subsection{Monte Carlo Markov Chain}

The \textit{uSHO} equation was fitted to ACFs using a Monte Carlo Markov Chain - MCMC. An MCMC is a means to `random walk' towards the and to sample the joint posterior probability distribution of the fitted parameters. By estimating initial values for the parameters, $X_\theta$, an initial fit of the \textit{uSHO} equation is done and the likelihood, $\mathcal{L}$, measured through 
\begin{equation}
\ln \mathcal{L} = -\frac{\chi^2}{2} - \sum_{i=1}^{N} \left( \ln\sigma_{y_i} \right) - \frac{N}{2}\ln(2\pi)
\end{equation}
where
\begin{equation}
\chi^2 =  \sum_{i=1}^{N} \left( \frac{y_i - \mu}{\sigma_{y_i}} \right)^2.
\end{equation}
where $N$ is the number of ACF points, $y_i$ the value of the ACF points with the error $\sigma_{y_i}$, $\mu$ is the model ACF point value that corresponds to $y_i$. As the ACFs are often more distorted from the \textit{uSHO} trend at higher time lags, due to interference from new starspots coming into effect, the MCMC only fits up to a time lag equivalent to $2.5\times P$.

The parameter values are then perturbed by a small amount to a new position in parameter space and the fit and likelihood calculations are repeated. If the likelihood is higher than the previous likelihood then the step is accepted and the next step takes place from the current location in parameter space. If the likelihood is worse than previous, it may be accepted under the Metropolis-Hastings algorithm \citep{Metropolis_et_al_1953, Hastings_1970}, otherwise it will be rejected and the step is not completed and it goes back to the previous step and randomly steps again.

The Metropolis-Hastings algorithm enables occasional steps in the wrong direction to ensure that an MCMC does not become trapped at a local likelihood maximum, and to enable exploration of the entire likelihood landscape. An optimum acceptance rate for an N-dimensional MCMC is approximately 0.25 \citep{Roberts_Gelman_Gilks_1997}. Rates much lower or higher than this may struggle to converge. To achieve this, an optimal step size is calculated from the curvature of the $\chi^2$-parameter space for each parameter $\alpha$,
\begin{equation}
\sigma_{X_i} = \sqrt{\frac{2}{\partial^2\chi^2/\partial^2\alpha^2}}\quad ,
\end{equation}
where the exact step size per MCMC step is a Gaussian distribution using $\sigma_{X_i}$ and centred on the previous parameter value.

The initial inputs of the parameters for the MCMC are estimated from the ACF or given standard values: period in days, determined as the time lag of the largest side lobe of the ACF, representative of the rotation period; the decay time \textit{$\tau_{AR}$} is based on the ratio of the first and second peaks of the ACF,
\begin{equation}
\tau_{AR} = -\frac{P} {\log\left(\frac{y_i(P)} {y_i(0)} \right)};
\end{equation}
\textit{A} is the ACF value at time lag = 0; and \textit{B} and $y_0$ are taken to be zero.

As a means to encourage the MCMC not to search for solutions in the unlikely areas of parameter space, Gaussian priors were applied to three of the parameters: amplitude \textit{A}, P and $\log\tau_{AR}$. For $\tau_{AR}$, having a Gaussian prior in log space reduces the risk of the MCMC wandering to unlikely high values. Also a hard lower limit of 1 day was included for $\log\tau_{AR}$ to prevent a highly improbable $\tau_{AR}$ value.

To determine whether convergence has been achieved, we adopt a likelihood rule as used by \citet{Charbonneau_et_al_2008} and \citet{Knutson_et_al_2008}. Each calculated likelihood $\mathcal{L}$ was stored and the current likelihood compared to the median of all those previous. When $\mathcal{L}$ falls below the median, the MCMC is considered to have achieved convergence. The MCMC then conducts another 5000 steps from which the mean and the standard deviation of each parameter are measured. This then launches a second MCMC routine using the mean and standard deviations as new initial parameters, $X_\theta$, and step sizes ($\pm \sigma_{X_\theta}$). This second MCMC explores the likelihood maximum to find the optimal parameter values. Two final tests for convergence are applied to the final 5000 steps of the second MCMC chain: we calculate the correlation length of this chain (and check that it is less than $\sim5\%$ of the total chain) and compute the Gelman-Rubin test \citep{Gelman_Rubin_1992}. Only stars that passed both of these tests are considered completed. These stars were then quickly visually inspected to remove any where the fits were obviously wrong.
Additionally, a check for correlations of all the fits of the ACFs for the targets was conducted by comparing all the parameter values to one another.

\begin{figure}
	\includegraphics[width=0.5\textwidth,keepaspectratio]{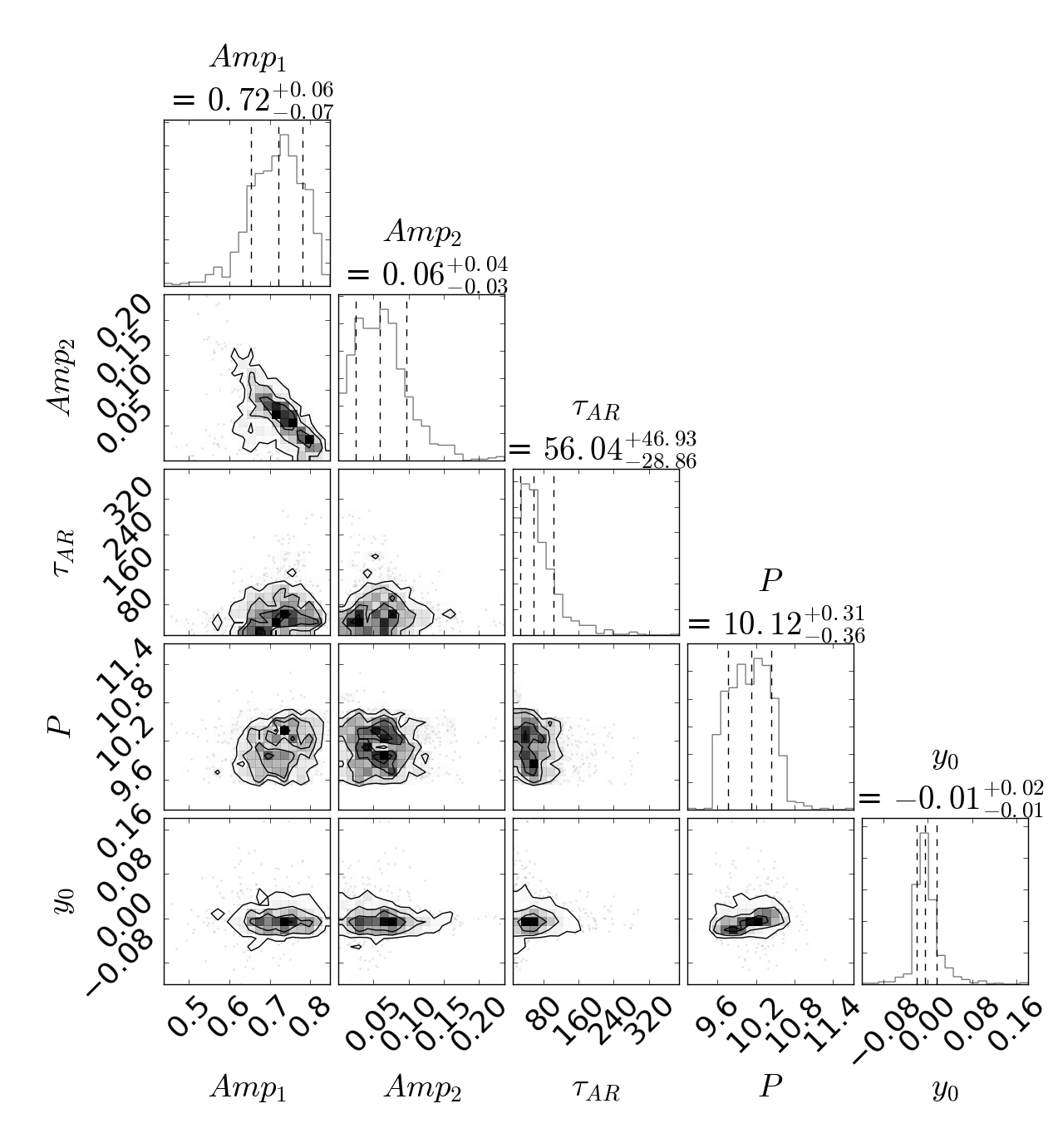}
    \caption{Correlation of all five MCMC parameters for the 10 day period sample. Most show Gaussian distributions apart from those associated with the offset -- they indicate that the offset value is dependent on other parameters. There is also a correlation between the two ACF amplitudes, which is not surprising as typically if there was an interpulse present in a target's ACF then the larger the interpulse, the smaller the primary amplitude.}
    \label{fig:corner_10d}
\end{figure}

\begin{figure}
	\includegraphics[width=0.5\textwidth,keepaspectratio]{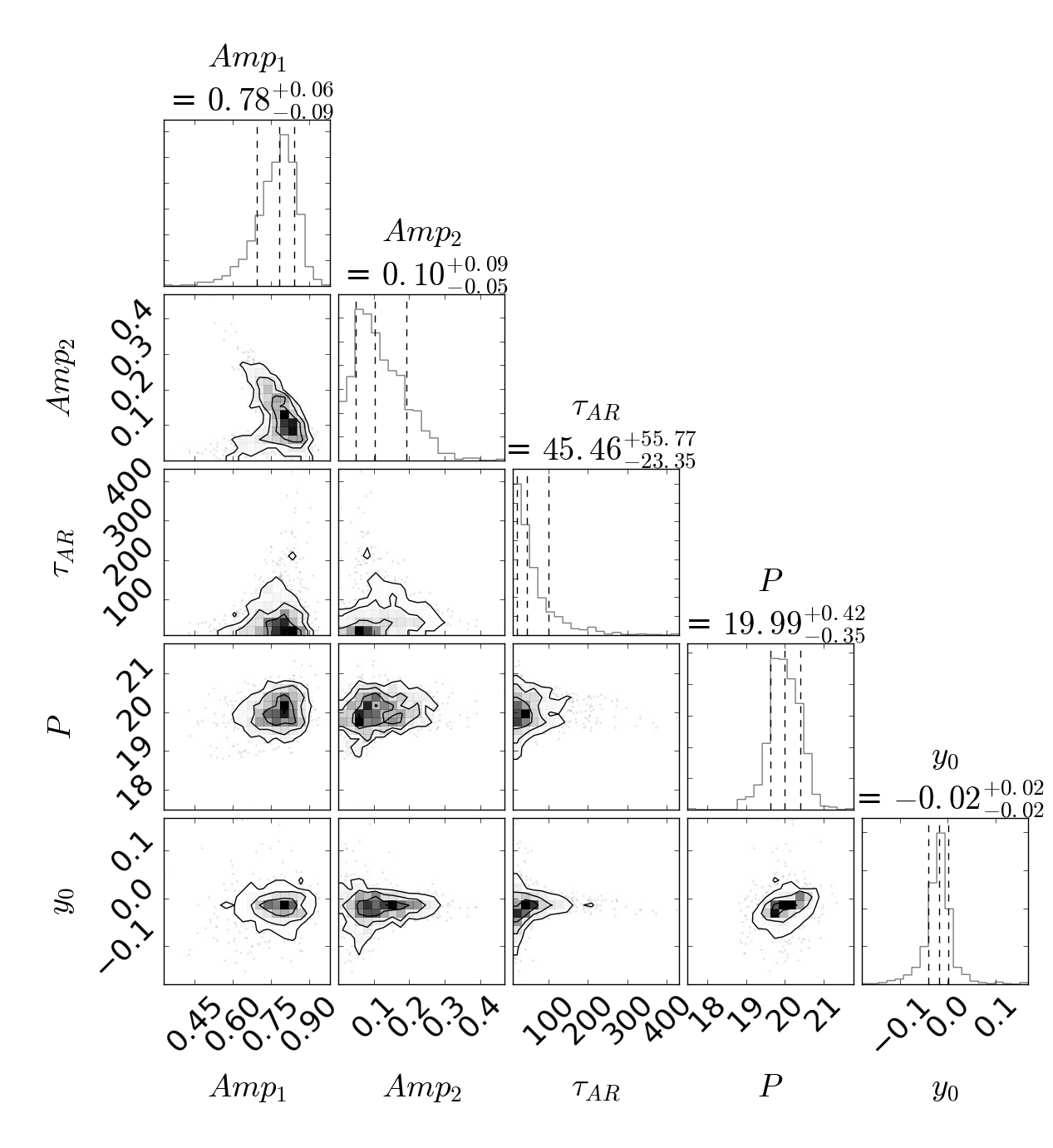}
    \caption{Correlation of MCMC parameters of the 20 day period sample. All have Gaussian distributions apart from the two ACF amplitudes, which typically have smaller primary amplitudes when the interpulse amplitude increases.}
    \label{fig:corner_20d}
\end{figure} 

In Figs. \ref{fig:corner_10d} and \ref{fig:corner_20d}, it can be seen that there are no strong unexpected correlations. The small correlation between the two amplitude sizes is not concerning as when there is an interpulse present in an ACF this reduces the initial amplitude at zero-time lag. Therefore, the larger the interpulse amplitude, the smaller the initial amplitude.

\subsection{\textit{Kepler} Light Curve Morphologies}
\label{sec:lc_morphologies}

\begin{figure*}
	\includegraphics[width=\textwidth]{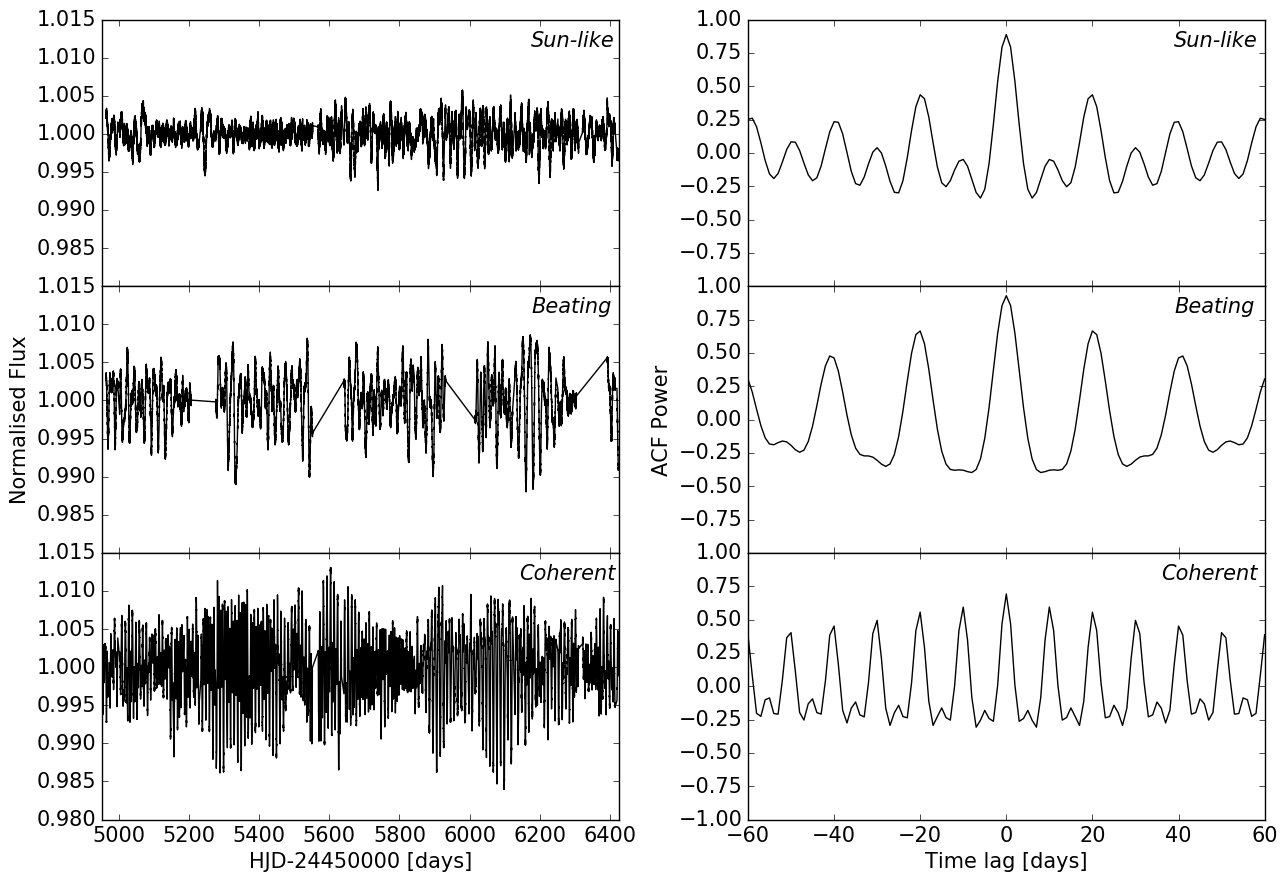}
    \caption{Three example light curves showing the three distinct light curve morphologies often seen in \textit{Kepler} data, and their auto-correlation functions. The Sun-like star, \textit{KIC 2985814}, shows starspots which have a decay lifetime similar to the rotational period. The beating star (``beater"), \textit{KIC 11802642}, has starspots which manage to survive a couple of stellar rotations and presents with a beating effect in the light curve. The long $\tau{\rm AR}$ star, \textit{KIC 8869186}, is very coherent and has starspots which last many rotations. All of these decay lifetimes can be quite easily seen in how the peaks decay away in the autocorrelation functions. Taking a ratio of the decay lifetime over the rotational period, each morphology can be defined as $\sim1$, $>1$ and $\gg1$ for Sun-like, `Beaters' and coherent stars respectively.}
    \label{fig:lightcurves_ACFs}
\end{figure*}

There are three distinct types of light curve morphologies (Fig.~\ref{fig:lightcurves_ACFs}) that can be seen in the bulk of \textit{Kepler} data - `Sun-like', `Beater' and `Coherent'. These are purely qualitative descriptions. On the other hand, inspecting the autocorrelation functions, a distinction can be seen. `Sun-like' stars appear to have starspot decay lifetimes that last approximately a rotational period, `Beaters' have lifetimes that last a few rotations and the `Coherent' stars have spots that persist for many rotations. Thereby taking the ratio of the activity starspot lifetime versus rotational period, ${\tau_{\rm AR}}/{P_{\rm rot}}$ (AR=Active Region, rot=rotation), we can define the ratio for each light curve morphology as $\sim1$ for Sun-like stars, $>1$ for `Beaters' and $\gg1$ for the `Coherent' stars.
It is known from Doppler imaging studies that many very active, fast-rotating stars have large, dark polar spots \citep[][and references therein]{Vogt_Penrod_1983,Strassmeier_2009}. Unless they are perfectly axisymmetric, such large polar features are likely to give rise to quasi-sinusoidal modulation. Since polar spots are generally large, we might expect them to have long lifetimes, producing modulations that would remain coherent for many rotation cycles. At the modest activity levels of most \textit{Kepler} stars, however, such large polar spots are not expected to be widespread.

\subsection{Determining the Starspot Sizes}

Whilst it is possible to determine approximate spot sizes for FGK-stars from Doppler imaging \citep{CollierCameron_1995, Barnes_James_CollierCameron_2002}, there is currently no direct method to measure them from light curves. However, light curves do have continuous variations -- these occur due to asymmetry between two sides of the star.
It is worth making the point that the amplitude of solar photometric variability increases with overall activity levels through the magnetic cycle \citep{Krivova_et_al_2003}. This implies that the power-law distribution of active-region sizes is such that the largest individual active regions dominate the modulation. If all active regions were of similar size, an increase in the number of active regions at different longitudes would cause the light curve modulation amplitude to decrease rather than increase.\citep{Bogdan_et_al_1988}
Therefore, as a proxy, the root-mean-square (RMS) scatter of the light curve can be extrapolated to be representative of starspot size.
\begin{equation}
RMS = \sqrt{ \frac{1}{N} \sum_{i=1}^{N}y_{i}^{2} }
\end{equation}
N is the total number of points in the light curve and $y_{i}$ the value of the flux at each data point. For a target, the 2-$\sigma$ range of the RMS (which encompasses $\sim$95\% of points) is calculated, as this encompasses the majority of the sinuous structure of the light curve but ought not include the erroneous outliers which may not have all been removed during post-observation processing.

\section{Results}
\label{sec:results}

\begin{figure*}
	\includegraphics[width=\textwidth]{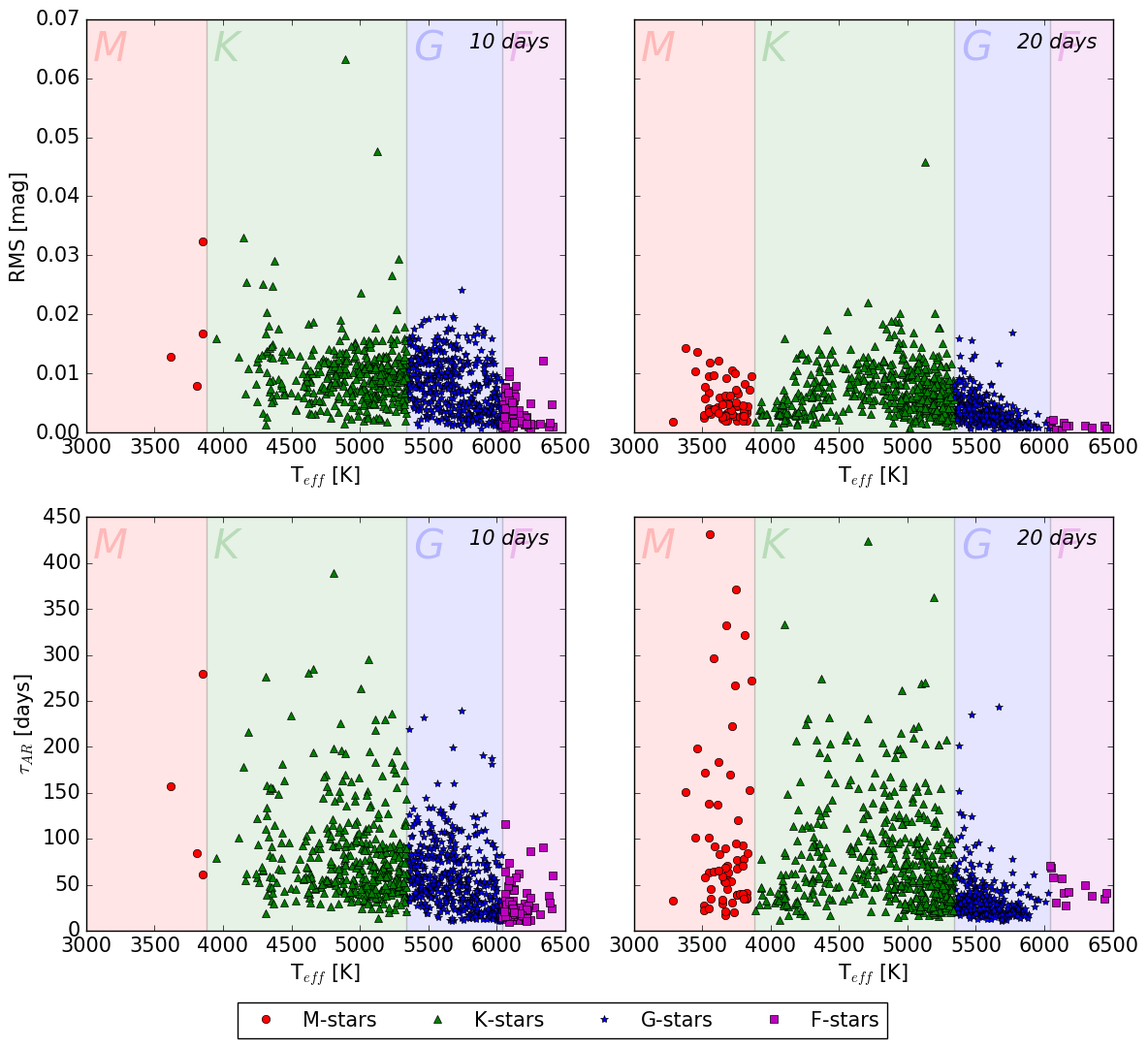}
    \caption{Plot series showing both data sets in two configurations. Upper level: effective temperature of targets (as stated by \citetalias{McQuillan_et_al_2014}) vs. the RMS of the targets' light curve. Lower level: effective temperature of targets vs. the measured decay lifetime. All targets have been split in colour and symbol based on their spectral type (from M- to F-stars) determined from \citet{Pecaut_Mamajek_2013}. For both data sets, the average spot size (RMS) and decay lifetime decrease the hotter the star.}
    \label{fig:four_plot_RMS_tau_Teff}
\end{figure*}

Generally the quality of the fits produced by the MCMC routine were good, though some were poorer and a couple were entirely spurious fits. Therefore all of the results were also inspected by eye and those with significantly different fits, therefore not representative, were rejected from the sample.

With 1089 stars for the 9.5-10.5 day (i.e. 10 day) period sample and 1154 stars for the 19.5-20.5 day (i.e. 20 day) period sample, the ACF fitting program returned 913 (83.8\% success rate) and 861 (74.6\% success rate) acceptable ACF fits for the 10 day and 20 day sample respectively.

In Fig. \ref{fig:four_plot_RMS_tau_Teff} the targets have been partitioned by spectral type (from M- to F-stars) as determined from \citet{Pecaut_Mamajek_2013}, and are represented by different colours and symbols which are detailed in the attached key. The first row shows the how the RMS amplitude of the rotational modulation (proxy for the starspot size for a star) varies with the stellar effective temperature for each of the two samples. The second row displays how the decay lifetime depends on the effective stellar temperature.

\subsection{Comparison of Rotation Periods}

\begin{figure*}
	\includegraphics[width=\textwidth]{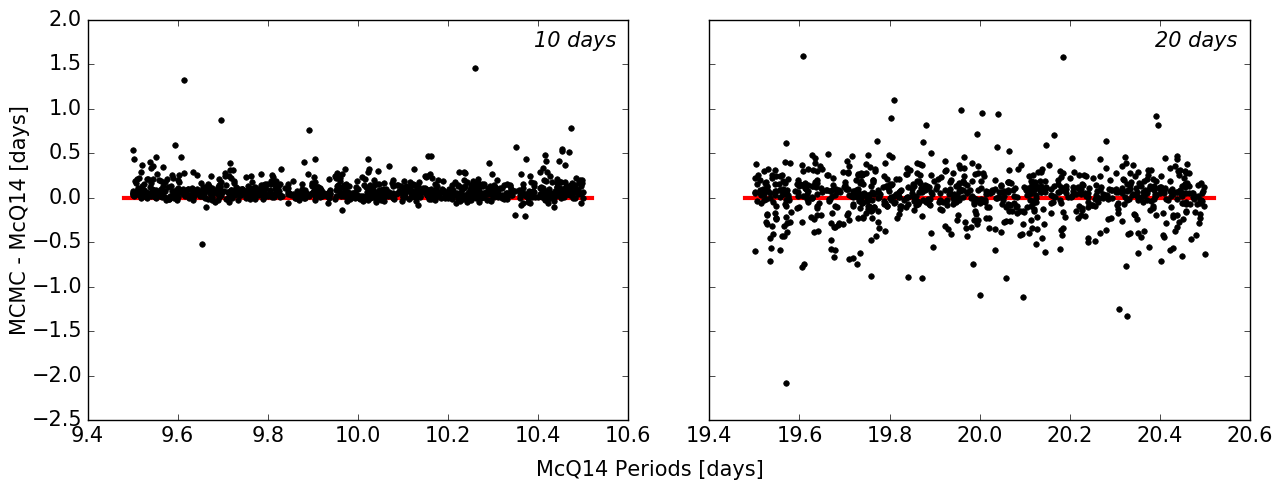}
    \caption{Comparison of the MCMC-measured stellar rotation period and the period determined by \citetalias{McQuillan_et_al_2014}. The red line represents the line where the MCMC-measured period is the same as those from \citetalias{McQuillan_et_al_2014}. For both the 10 day period sample, and in particular the 20 day period sample, there is a large range of differences in periods. However, something to note is the difference in auto-correlation function generation from \citetalias{McQuillan_et_al_2014} and that an MCMC was then applied to the different ACF.}
    \label{fig:duo_period_comparison}
\end{figure*}
\begin{figure*}
	\includegraphics[width=\textwidth]{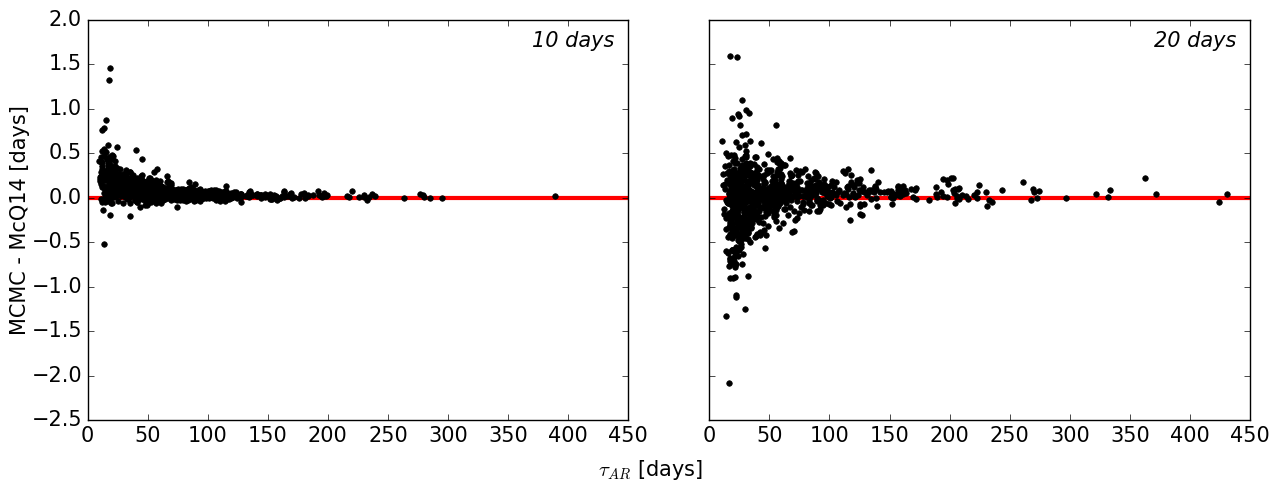}
    \caption{Comparison of the MCMC-measured stellar rotation period and the period determined by \citetalias{McQuillan_et_al_2014} with respect to $\tau{\rm AR}$. The red line indicates where the MCMC-measured period is the same as \citetalias{McQuillan_et_al_2014}. The 10 day period sample shows an asymmetry in the residuals indicating that for the smaller decay lifetimes $\tau_{\rm AR}$ there is a larger disagreement between the two measured periods. This is most likely due to \citetalias{McQuillan_et_al_2014} underestimating the true period as they did not consider the decay envelope. In the 20-day sample, short active-region lifetimes degrade the precision with which the rotation periods can be determined, leading to a more symmetric distribution in the differences between periods determined with the two methods.}
    \label{fig:duo_period_comparison_tau}
\end{figure*}

In \citetalias{McQuillan_et_al_2014} the periods were determined using an autocorrelation function routine, and these were used during sample selection. Comparing the periods from \citetalias{McQuillan_et_al_2014} and those generated by the MCMC (Fig.~\ref{fig:duo_period_comparison}), there is some variation with the 10 day sample varying less than the 20 day sample. This range will reflect upon the difference in autocorrelation function generation as the routines used in \citetalias{McQuillan_et_al_2014} and this paper are different, meaning variation in stellar rotations periods is to be expected. Further, as a point of interest, the residuals for the 10 day sample are asymmetric, with our algorithm generally finding longer periods than \citetalias{McQuillan_et_al_2014}. Due to not fitting the decay envelope, \citetalias{McQuillan_et_al_2014} will have underestimated the period, biasing the first sidelobe to a lower time lag. Therefore, the shorter the decay lifetime, the larger a discrepancy seen in Fig.~\ref{fig:duo_period_comparison_tau}. Interestingly, this becomes symmetric for the 20 day sample, but with the same trend that shorted decay lifetimes have larger range.

\subsection{10 Day Period Sample}

For this sample, in Fig. \ref{fig:four_plot_RMS_tau_Teff} (left-hand side), there is a distinct distribution of starspot sizes and decay lifetimes. Hotter stars with T$_{\rm eff}$ greater than 6200K, have a smaller range of spot sizes than cooler stars. These stars also have spots which do not survive for very long. At effective temperatures above the $\sim6200$K boundary, the limit on decay lifetime is less than 100 days. This is up to a third of starspot lifetimes on much cooler stars.

\begin{figure*}
	\includegraphics[width=\textwidth]{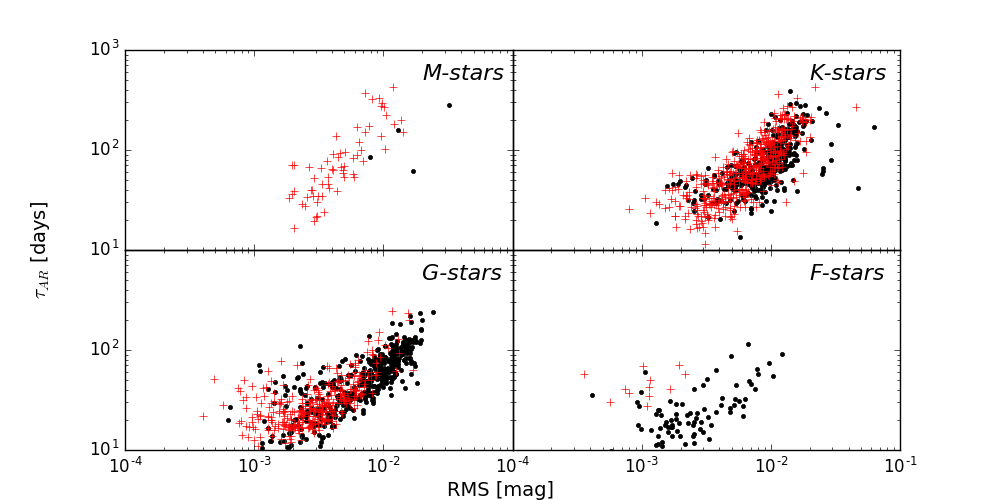}
    \caption{Distribution of decay timescales and RMS of target light curves, split by spectral type (based on temperature boundaries from \citet{Pecaut_Mamajek_2013}) for the 10 ($\bullet$) and 20 (\textcolor{red}{+}) day period sample. There is a slight increase in trend gradient as stellar temperature increases. There is a strong relationship between the day lifetime and RMS -- the larger the RMS of the light curve, the larger the decay lifetime. For the hottest stars, the size of spots possible appear to be very small, and they often do not survive very long.}
    \label{fig:quad_RMS_tau}
\end{figure*}

For ease of viewing, the comparison between spot size and decay lifetime has been split into the four observed spectral types in Fig. \ref{fig:quad_RMS_tau}. The coolest stars (M-stars) have a large range of spot size vs. decay timescale but given the very small stellar population this is not representative. However, there are a great many more K-stars and G-stars which show a strong trend of longer decay lifetimes for larger spots. The gradient of the trend is greater for the K-stars, indicating that the hotter the star, the shorter the lifetime. Additionally, the range of the spot sizes associated with the G-stars is less than the K-stars. This limits spots to have no larger effect on the light curve than an RMS of 0.025 mag. The F-stars, like the M-stars are not very numerous in this sample. However, they do all cluster together at low decay lifetimes and small spot sizes suggesting that for this the hottest of all the targets, spots rarely reach a large size or survive very long. This would also suggest spots survive longer the bigger they are.

\subsection{20 Day Period Sample}

The 20 day sample is similar to the previous sample with a few small differences (Fig.~\ref{fig:four_plot_RMS_tau_Teff}): the temperature above which the range of spot sizes dramatically decreases is at a lower temperature $\sim$5700K and spots can survive longer on cooler stars than in the 10 day sample.

As for the 10-day sample, when we partition the stars by spectral type for the relationship between decay lifetime and spot size (Fig. \ref{fig:quad_RMS_tau}), the coolest stars again are not well represented. For the K- and G-stars there is again a positive relationship with increasing decay lifetime and larger spots, with the trend gradient appearing to just be slightly steeper for the K-stars. However, the range of decay lifetimes and spot sizes is much more limited for these G-stars than in the other sample. The F-stars similarly cluster in the lower decay lifetime, smaller spot size area, but have a little more range than the 10 day period sample of F-stars.

\subsubsection{Solar Comparison}
\label{sec:solar}

From investigations on stars observed by \textit{Kepler} and previous surveys, there was discussion about the activity of the Sun and whether it was unusually quiet \citep{Radick_et_al_1998,Gilliland_et_al_2011}. Comparing it to the 20 day sample (solar rotation period $\sim27$ days), stars with Sun-like temperatures ($\sim5800$K) all have small light curve amplitudes indicating small spots.
The amplitudes of solar variability measured by \citet{Krivova_et_al_2003} through the solar cycle are very similar to those measured in this work for stars with solar-like rotation periods and effective temperatures.
This would \citep[as discussed in][]{Basri_Walkowicz_Reiners_2013} indicate that the Sun is not suspiciously inactive.

\subsection{Spot Size and Distribution}

\subsubsection{RMS as a Proxy for Spot Size}
We find that stars with large RMS-variations indicate spots with longer lifetimes. This could lead to two interpretations: large variations could mean that there are a few big spots dominating with smaller RMS variations meaning there are only small spots. But it could theoretically be possible that there are many spots of a similar size. There is good physical reasoning behind the hypothesis that diffusive decay takes longer to destroy big active regions than small ones. If indeed the lifetime is short for stars that have many spots of similar size, short lifetimes would also be associated with small light curve amplitudes. Implementing Occam's Razor, the simpler explanation is, however, that the solar spot-size and spot-lifetime power laws can be extrapolated to other stars, and that the same physical processes operate.

\subsubsection{Active-Region Lifetime as a Function of Spot Size and Effective Temperature}

Using the two datasets together, it is possible to generate a function using the RMS (as a spot size proxy) and effective temperature to generate an expected active region lifetime which can be used for an individual star. Orthogonalising the data by removing the mean value of each distribution and fitting a quadratic through regression to the data in log-log space, the following relation is determined:
\begin{equation}
\begin{split}
\log_{10}{\tau_{AR}} = 10.9252 + 3.0123\cdot\log_{10}{\rm RMS}\: + \\
+\: 0.5062\cdot\left(\log_{10}\rm RMS\right)^{2}- 1.3606\cdot\log_{10}{T_{\rm eff}}
\end{split}
\end{equation}
where \textit{RMS} is the RMS scatter of individual \textit{Kepler} light curves which were normalised to a mean flux of unity, \textit{T$_{\rm eff}$} is the stellar effective temperature in K, and \textit{$\tau_{\rm AR}$} is the resultant decay lifetime in days. If this is used as an estimate for the mean of a Gaussian prior probability distribution for $\log{\tau_{\rm AR}}$ then the standard deviation $\sigma$ of the residuals from the fit should be used as the standard deviation $\sigma$ of the prior: $\sigma \left( \log_{10} \tau_{\rm AR} \right) = 0.178623$.

\subsubsection{Active Longitudes}
When considering active longitudes, evidence from the \textit{Kepler} light curves suggests that even if spots persistently recur at active longitudes, they would tend to preserve the coherence of the light curve on timescales longer than the lifetimes of an individual active region. We cannot explicitly say whether such an effect is present, however we note that the decay timescales we obtain from the light curves of the solar-like stars are comparable with the lifetimes of the large solar spot groups.

\section{Conclusion}
\label{sec:conclusions}

The subject of this paper was to determine whether there is a relationship between the sinusoidal amplitude seen in \textit{Kepler} light curves, as a proxy for starspot size, and the decay timescale of starspots. Furthermore, we sought to determine whether the lifetimes of spots of a given size depend on the stellar effective temperature.

As seen within the two samples (9.5-10.5 days and 19.5-20.5 days period stars) drawn from \citetalias{McQuillan_et_al_2014}, there are three main conclusions:
\begin{enumerate}
\item Big starspots live longer on any given star,
\item Starspots decay more slowly on cooler stars and
\item The Sun is not unusually quiet for its spectral type.
\end{enumerate}

Our observation that big spots generally survive longer longer on any given star is consistent with models of spot decay in which turbulent diffusion is eating the edges of the spots \citep{Simon_Leighton_1964,Litvinenko_Wheatland_2015,Litvinenko_Wheatland_2016}. This is also consistent with our finding that spots generally survive longer on cooler stars. As the vigour of convection is temperature dependent, the turbulent diffusivity, and hence the rate of spot decay, will increase with the convective heat flux and hence with effective temperature. An analogy would be food colouring being dispersed more slowly in cool water than in boiling water.

The work presented in this paper has deepened our knowledge of the connection between the light curve morphologies of Kepler stars and the physics that determine active-region lifetimes in convective stellar photospheres. This in turn can be applied to many areas which rely on light from stars, in particular when searching and analysing exoplanet host candidates.

\section*{Acknowledgements}

We would like to thank our referee for their constructive comments that have improved the quality of this work.

This paper includes data collected by the \textit{Kepler} mission. Funding for the \textit{Kepler} mission is provided by the NASA Science Mission directorate. This research has made use of the NASA Exoplanet Archive, which is operated by the California Institute of Technology, under contract with the National Aeronautics and Space Administration under the Exoplanet Exploration Program.

HACG acknowledges the financial support of the National Centre for Competence in Research `PlanetS' supported by the
Swiss National Science Foundation (SNSF) and the financial support of the University of St Andrews; and the computer support from the University of St Andrews. ACC acknowledges support from STFC consolidated grant number ST/M001296/1. RDH gratefully acknowledges support from STFC studentship grant ST/J500744/1, a grant from the John Templeton Foundation, and NASA XRP grant NNX15AC90G. The opinions expressed in this publication are those of the authors and do not necessarily reflect the views of the John Templeton Foundation.
This research was submitted as a Masters project for HACG supervised by ACC at the University of St Andrews in April 2015. HACG gratefully thanks for all supervision from ACC and support from RDH and other members of the astronomy group.








%
%


\bsp	
\label{lastpage}
\end{document}